\documentclass[12pt, preprint]{aastex}

\begin{document}

\title{The Interplanetary Network Supplement to the BATSE Catalogs of Untriggered Cosmic
Gamma-Ray Bursts}

\author{K. Hurley\altaffilmark{1}}

\email{khurley@ssl.berkeley.edu}

\author{B. Stern\altaffilmark{2,3,4}}

\author{J. Kommers\altaffilmark{5}}

\author{T. Cline\altaffilmark{6}}

\author{E. Mazets, S. Golenetskii\altaffilmark{7}}

\author{J. Trombka, T. McClanahan\altaffilmark{8}}

\author{J. Goldsten\altaffilmark{9}}

\author{M. Feroci\altaffilmark{10}}

\author{F. Frontera\altaffilmark{11,12}}

\author{C. Guidorzi\altaffilmark{13,12}}

\author{E. Montanari\altaffilmark{12}}

\author{W. Lewin\altaffilmark{14}}

\author{C. Meegan, G. Fishman, C. Kouveliotou\altaffilmark{15}}

\author{S. Sinha, S. Seetha, \altaffilmark{16}}

\altaffiltext{1}{University of California, Berkeley, Space Sciences Laboratory,
Berkeley, CA 94720-7450}

\altaffiltext{2}{Institute for Nuclear Research, Russian Academy of Sciences,
Moscow 117312, Russia.}

\altaffiltext{3}{Astro Space Center of the Lebedev Physical Institute,
Profsoyuznaya 84/32, Moscow 117810, Russia.}

\altaffiltext{4}{SCFAB, Stockholm Observatory, SE-10691, Stockholm, Sweden}

\altaffiltext{5}{MIT Lincoln Laboratory, Lexington, MA 02420-9108}

\altaffiltext{6}{NASA Goddard Space Flight Center, Code 661, Greenbelt, MD 20771}

\altaffiltext{7}{Ioffe Physico-Technical Institute, St. Petersburg, 194021, Russia}

\altaffiltext{8}{NASA Goddard Space Flight Center, Code 691, Greenbelt, MD 20771}

\altaffiltext{9}{The Johns Hopkins University, Applied Physics Laboratory, Laurel, MD 20723}

\altaffiltext{10}{Istituto di Astrofisica Spaziale, C.N.R.,
  Area di Ricerca Tor Vergata, Via Fosso del Cavaliere 100,
 00133 Roma, Italy}

\altaffiltext{11}{Istituto Astrofisica Spaziale e Fisica Cosmica, C.N.R., Sezione di Bologna, Via
Gobetti 101, 40129 Bologna, Italy}

\altaffiltext{12}{Dipartimento di Fisica, Universita di Ferrara, Via Paradiso 12,
44100, Ferrara, Italy}

\altaffiltext{13}{Astrophysics Research Institute, Liverpool John Moores University,
Twelve Quays House, Egerton Wharf, Birkenhead CH41 1LD, United Kingdom}

\altaffiltext{14}{Massachusetts Institute of Technology, Center for Space Research 37-627, 
Cambridge MA 02139}

\altaffiltext{15}{NASA/MSFC, National Space Science and Technology Center, SD-50,
320 Sparkman Drive, Huntsville, AL 35805}

\altaffiltext{16}{ISRO Satellite Center, Space Astronomy and Instrumentation Division,
Bangalore 560 017, India}

\begin{abstract}

We present Interplanetary Network (IPN) detection and localization information for 211 
gamma-ray bursts (GRBs) observed as untriggered events by the Burst and Transient Source Experiment (BATSE),
and published in catalogs by Kommers et al. (2001) and Stern et al. (2001).  IPN confirmations have been 
obtained by analyzing the data from 11 experiments.  For any given
burst observed by BATSE and one other distant spacecraft, arrival time analysis
(or ``triangulation'') results in an annulus of possible arrival
directions whose half-width varies between 14 arcseconds and 5.6 degrees, depending
on the intensity, time history, and arrival direction of the burst, 
as well as the distance between the spacecraft.  This annulus generally
intersects the BATSE error circle, resulting in a reduction of the
area of up to a factor of $\sim$ 650.  
When three widely separated spacecraft observed a burst, the result is an error
box whose area is as much as 30000 times smaller than that of
the BATSE error circle. 

Because the IPN instruments are considerably less sensitive than
BATSE, they generally did not detect the weakest untriggered bursts, but did detect the more intense ones which
failed to trigger BATSE when the trigger was disabled.  In a few cases, we
have been able to identify the probable origin of bursts as soft gamma repeaters.
The vast majority of the IPN-detected events, however, are GRBs, and the confirmation of them
validates many of the procedures utilized to detect BATSE untriggered bursts.

\end{abstract}


\section{Introduction}

This paper presents the 8th catalog of gamma-ray burst (GRB) localizations 
obtained by arrival time analysis, or ``triangulation'' between the 
missions in the 3rd interplanetary network (IPN), which began operations in
1990 and continues to operate today.
Two of these catalogs (Hurley et al. 1999a,b) were supplements to the BATSE
3B and 4Br burst catalogs (Meegan et al. 1996; Paciesas et al. 1999).  The others
involved bursts observed by numerous other spacecraft (Laros et al. 1997, 1998;
Hurley et al., 2000a,b,c).  In
this paper, we present IPN data on 211 \it untriggered \rm bursts which occurred throughout the entire 
\it Compton Gamma-Ray Observatory (CGRO) \rm mission (1991 April through
2000 May).  The BATSE data on these events, such as durations, fluxes, fluences, and
coarse location information, appear in two catalogs, 
Kommers et al. (2001) and Stern et al. (2001).  A final IPN supplement catalog, to the
BATSE 5B catalog, is in preparation (Hurley et al. 2004, Briggs et al., 2004).

The purpose of searching the BATSE data for untriggered events was mainly
to extend the number-intensity (log N-log S) distribution to weaker bursts than
those that could trigger the detector, and thus to gain more information on the burst
distribution, particularly at the weak end.  Other objectives included the
detection of bursts from known and unknown soft gamma repeaters, and very
soft transients which could constitute a previously unknown phenomenon.  (One significant
outcome of this effort was the detection of the bursting pulsar).  
The purpose of searching the IPN data for these events was to
confirm as many of them as possible, reduce the sizes of their error circles, and validate the procedures
used to identify these untriggered events.

\section{Instrumentation, Search Procedure, Derivation of Annuli, and
Burst Selection Criteria}

We have used the same procedures as those employed in the other BATSE
catalog supplements, and refer the reader to Hurley et al. (1999a,b) for the detailed descriptions.
Generally speaking, using the arrival time and direction of a burst at BATSE, and
its time history, we searched
the data of the near-Earth spacecraft for a confirmation at the same time; for the spacecraft
which were far from Earth, we searched for a confirmation (i.e., an event with a matching
time history) 
in the appropriate crossing time window.  Although  more than 15 separate gamma-ray burst
experiments were operating on over a dozen missions throughout the duration of the CGRO mission, 
confirmations were obtained from the data of just
11 experiments: the \it BeppoSAX \rm Gamma-Ray Burst Monitor (Frontera et al. 1997; Feroci et al. 1997), 
the \it Defense Meteorological Satellite Program \rm (DMSP, Terrell et al. 1992), 
Ginga (Murakami et al. 1989), Konus-A (Aptekar et al.
1997), Konus -\it Wind \rm (Aptekar et al. 1992), the Near Earth Asteroid Rendezvous mission (NEAR,
Goldsten et al. 1997), PHEBUS (Terekhov et al. 1994), \it Pioneer
Venus Orbiter \rm (Klebesadel et al. 1980), SROSS C-2 (Kasturirangan et al. 1997),
\it Ulysses \rm (Hurley et al. 1992), and
WATCH - GRANAT (Brandt, Lund, \& Rao 1990). 
We note here, however, two important differences in the procedures and
results between the triggered and untriggered events.

First, the
untriggered burst catalogs contain a much higher proportion of weak events than the
BATSE triggered burst catalogs.  Because
the IPN instruments are generally much less sensitive than BATSE, they detected a
smaller fraction of the untriggered than the triggered ones.

Second, the untriggered event time histories were recorded in the 1.024 s resolution BATSE data,
while the triggered event time histories were recorded with much higher time resolution.
Thus when an untriggered event was detected only by BATSE and another
near-Earth spacecraft, the low time resolution and the proximity of the two spacecraft results in a very wide annulus
which is consistent with, but does not constrain the BATSE error circle.  Twenty-one events
fell into this category, and it is only possible to confirm their detection,
but not to obtain a meaningful annulus or error box for them.

\section{A Few Statistics}

There are 873 untriggered bursts in the Kommers et al. (2001) catalog and 1838 
untriggered bursts in the Stern 
et al. (2001) catalog.
The two sets are not mutually exclusive (Stern et al. 2001), and the total
number of untriggered bursts is approximately 2000, depending on the 
exact acceptance criteria.  Their peak fluxes range from 0.06 to 25 photons cm$^{-2}$ s$^{-1}$.
Figure 1 gives the IPN efficiency for detecting untriggered bursts as a function
of their peak fluxes.  This is defined as the number of bursts detected by
the IPN divided by the total number of untriggered bursts in a particular flux
range.  There are many factors which determine whether a burst is detected by an IPN
spacecraft.  In addition to the burst intensity and time history, solar activity, Earth-blocking
for spacecraft in low Earth orbit, the number of spacecraft active in the IPN, and data return 
all play important, time-variable roles.  Figure 1 therefore gives time-averaged efficiencies.
Approximately one out of nine untriggered BATSE bursts was observed by at least one spacecraft
in the IPN.  Their fluxes range from 0.15 to 25 photons cm$^{-2}$ s$^{-1}$.
For comparison, approximately one out of every three triggered BATSE bursts
was observed by IPN spacecraft (Hurley et al. 1999b).  Of the 211 IPN events,
only 90 could be localized (85 to annuli only, and 5 to error boxes).

\section{Tables of Confirmed Bursts, Annuli, and Error Boxes}

For each confirmed untriggered burst, table 1 lists the spacecraft which observed the event.
(A list of \it all \rm GRBs and the IPN spacecraft which detected them may be found
at \url{http://ssl.berkeley.edu/ipn3/masterli.html} or \url{http://heasarc.gsfc.nasa.gov/W3Browse/}.  )

For those bursts which can be localized, either to a single annulus whose width is
comparable to or less than the diameter of the BATSE error circle (an example is
shown in figure 2), or to an error box (an example is shown in figure 3), the 6 columns in table 2 give:

1) the date of the burst, in yymmdd format, 
2) the Universal Time of the burst at Earth in seconds,
3) the right ascension of the center of the IPN 
annulus, epoch J2000, in the heliocentric frame, in degrees,
4) the declination of the center of the IPN 
annulus, epoch J2000, in the heliocentric frame, in degrees, 
5) the angular radius R$_{\rm IPN1}$ of the first IPN annulus, in the heliocentric
frame, in degrees, and
6) the half-width $\delta$ R$_{IPN1}$ of the first IPN annulus, in degrees; the 3 $\sigma$
confidence annulus is given by R$_{\rm IPN1}$ $\pm$ $\delta$ R$_{IPN1}$.

If the burst was detected by a third, distant spacecraft, and a non-degenerate
second annulus could be derived for it, the information in columns 4, 5, and 6 is
repeated for this annulus.

For the bursts in table 2, table 3 gives the BATSE error circles, from Kommers
et al. (2001) and Stern et al. (2001), and
either a) the intersection points of the IPN annulus with the error circle, or
b) for the three-spacecraft localizations, the four corners of the IPN error box.

For each entry, the first line contains:

1) the date of the burst, in yymmdd format, 
2) the Universal Time of the burst at Earth, in seconds,
3) the right ascension of the center of the BATSE error circle, in degrees,  
4) the declination of the center of the BATSE error circle, in degrees, and
5) the radius of the BATSE error circle, in degrees; this is the combination of the one sigma
statistical error and a 1.6 degree systematic error, summed in quadrature.

The four following lines contain the right ascension and declination, in degrees, of the error
box.  For those bursts which were observed by BATSE and a single IPN spacecraft (e.g.
figure 2),
the coordinates are those of the intersection of the 3 $\sigma$ IPN annulus with
the 1 $\sigma$ (statistical plus systematic) BATSE error circle.  Although all of
the annuli are statistically consistent with the positions of their respective 1 $\sigma$ BATSE error circles,
in some cases part or all of the annulus does not actually intersect the error circle.  
In those cases, the coordinates are set to zero.  For those bursts which were observed
by two distant IPN spacecraft (e.g. figure 2), and for which an IPN-only error box can be
derived, the coordinates given are those of the IPN error box.

All coordinates are J2000, and all event times are the ones used to identify the
bursts in the Stern et al. (2001) and Kommers et al. (2001) catalogs.

\section{Notes on specific events}

We note here a number of unusual circumstances surrounding some of the bursts
in the Stern et al. (2001) and Kommers et al. (2001) catalogs.

\begin{itemize}

\item Some of the bursts in the two catalogs
in fact correspond to BATSE triggers.  In some cases, the triggers were not caused
by the bursts, but the bursts were nevertheless recorded in triggered mode.

\item The Kommers et al. (2001) catalog was divided into two parts: high energy (HE) events, and
low energy (LE) events.  Initially, there were 125 LE events, but 75 of them were intentionally eliminated from the final
catalog because their origin was suspected to be either magnetospheric, X-ray binaries in
outburst, or activity from soft gamma repeaters (SGRs).  We have identified four of the 75 eliminated events as 
bursts from SGR1806-20.  These four can be found in the complete list of SGR bursts identified in the
Kommers et al. (2001) search, available at \url{http://space.mit.edu/BATSE/data.html}.

\item A total of 9 of the untriggered events probably originated from soft gamma
repeaters.  In some cases, they had in fact triggered BATSE and were recorded
in triggered mode.  The IPN localizations of these SGR bursts serve as a good calibration
of the techniques and data used here, however.  They verify, for example, that the 1 s resolution BATSE
data files in the Stern et al. (2001) and Kommers et al. (2001) catalogs have the correct timing,
and that the localization procedures used in these catalogs, and by the IPN, are accurate.

\end{itemize}

The following list gives the details.

GRB920903, 05728 s.  This burst was observed by WATCH-GRANAT (Sazonov et al. 1995).
Both the Ulysses/WATCH and Ulysses/BATSE annuli are consistent with the
WATCH error circle, but do not intersect the BATSE error circle.
The BATSE error circle lies $\sim$ 7 degrees away from the WATCH error
circle, but is consistent with it, given the statistical and systematic
uncertainties.  The intersection of the narrower Ulysses/BATSE 
annulus with the WATCH error circle is given here.

GRB920920, 04415 s.  This event was recorded in triggered mode following
BATSE trigger 1948.  The trigger occurred due to a different GRB.

GRB930209, 15737 s.  This event was recorded in triggered mode following 
BATSE trigger 2177.  The trigger occurred due to a solar flare.

GRB930702, 68333 s.  This burst corresponds to BATSE
trigger 2426, which is believed to be a Cygnus X-1 fluctuation.

GRB931005, 82288 s.  This burst is a Kommers et al. (2001) LE event which
was eliminated from the final catalog.  It is BATSE trigger
2565, from SGR1806-20.

GRB940710, 35477 s.  This event occurred
251 s before BATSE trigger 3071, whose duration is T90=71 s.
The location of the Kommers et al. (2001) event is RA, Decl. = 99.4$\degr$, 
-33.3$\degr$, with uncertainty
3.4$\degr$, and that of BATSE 3071 is RA, Decl. =  96.42$\degr$, -36.59$\degr$, with
uncertainty 1.33$\degr$.  The centers of the two circles are
therefore 4.1$\degr$ apart.  The IPN annulus is consistent
with both error circles.  Thus the Kommers et al. (2001) event may be a precursor
to 3071.

GRB950730, 76147 s.  This event corresponds to BATSE trigger 3720, which was  
due to a Cygnus X-1 fluctuation.  The burst occurred 25 s after the trigger.

GRB950904, 52777 s.  This event occurred about 75
s after BATSE trigger 3776.  Its duration is given as 108 s in Stern
et al. (2001), but this duration included the triggered event, which is
unrelated to it.  The correct duration is $\sim$ 30 s.

GRB961119, 21322 s.  This burst is a Kommers et al. (2001) LE event which
was eliminated from the final catalog.  The IPN annulus is consistent with
the position of SGR1806-20.

GRB961119, 21536 s.  This burst is a Kommers et al. (2001) LE event which
was eliminated from the final catalog.  It is probably from
SGR1806-20, but it cannot be triangulated with any precision.

GRB961119, 26961 s.  This burst is a Kommers et al. (2001) LE event which
was eliminated from the final catalog.  It is probably from
SGR1806-20, but it cannot be triangulated with any precision.

GRB980622, 51085 s.  This event is BATSE trigger 6861.  It originates from
SGR1627-41.

GRB980728, 64911 s.  This event is BATSE trigger 6954.  It originates from
SGR1627-41.

GRB980801, 12920 s.  This event is BATSE trigger 6959.  It originates from
SGR1627-41.

GRB980913 at 19983 s.  This is BATSE trigger 7087.

GRB981022, 56447 s.  This event is BATSE trigger 7171, from SGR1900+14.

GRB990110, 31141 s.  This is BATSE trigger 7315.  This burst originates
from SGR1900+14.

GRB990429, 35555 s.  This event is BATSE trigger 7536.  It originates from
SGR1900+14.

GRB990507, 71334 s.  This is BATSE trigger 7552.

GRB991101, 54480 s.  This is BATSE trigger 7835.  It is probably
a GRB observed with particle contamination.

\section{Discussion and Conclusion}

The Kommers et al. (2001) and Stern et al. (2001) studies of untriggered BATSE
bursts pointed to different conclusions about the GRB population.  The sample
of Stern et al. provides evidence for a GRB number-intensity relation which continues
to increase at low intensities, while the sample of Kommers et al. provides evidence
for a flattening.  The analysis which we have presented here indicates only that
many of the events with peak fluxes above $\sim$0.15 photon cm$^{-2}$ s$^{-1}$ are likely to be
real, and that relatively few of them have been misclassified.  The likelihood 
of reality increases with peak flux (figure 1).  As there are hundreds
of untriggered bursts below the IPN threshold, the possibility exists that the
different conclusions about the number-intensity relation are due to the
differences in classifying weak untriggered events.  A recent study of untriggered
BATSE bursts by Mitrofanov et al. (2004) reinforces and quantifies this idea.  While
this study is a preliminary one and does not draw any conclusions about the
weak events, it should eventually lead to a clearer classification of them.  A
definitive statement about the weak burst population may also be forthcoming after
the launch of the Swift mission (Gehrels et al. 2004).

\section{Acknowledgments}

Support for the \it Ulysses \rm GRB experiment is provided by JPL Contract 958056.  Joint
analysis of \it Ulysses \rm and BATSE data was supported by NASA Grants NAG 5-1560 and NAG5-9701.  NEAR
data analysis was supported under NASA Grants NAG 5-3500 and NAG 5-9503.  We are also
grateful to the NEAR team for their modifications to the XGRS experiment which made
gamma-ray burst detection possible.  The Konus-Wind experiment was supported by Russian
Space Agency Contract and RBRF grant \#03-02-17517.

\clearpage

FIGURE CAPTIONS

\figcaption{
The IPN efficiency for detecting a BATSE untriggered burst.  This is the number of
bursts in a flux range detected by the IPN, divided by the number detected by
BATSE.  The peak fluxes of the untriggered bursts range from 0.06 to 25 photons cm$^{-2}$ s$^{-1}$.
The efficiencies are time-averaged.\label{Fig. 1}
}

\figcaption{
The BATSE 1$\sigma$ (statistical + systematic) error circle for
the untriggered event on 980629, and the 3$\sigma$ IPN annulus. Note that in general the
curvature of the annulus makes it impossible to describe the resulting
error box with only the four annulus/error circle intersection points.\label{Fig. 2}
}

\figcaption{
The BATSE 1$\sigma$ (statistical + systematic) error circle for
the untriggered event on 000403, and the 3$\sigma$ IPN error box,
formed by the intersection of the two annuli. \label{Fig. 3}
}

\clearpage

\plotone{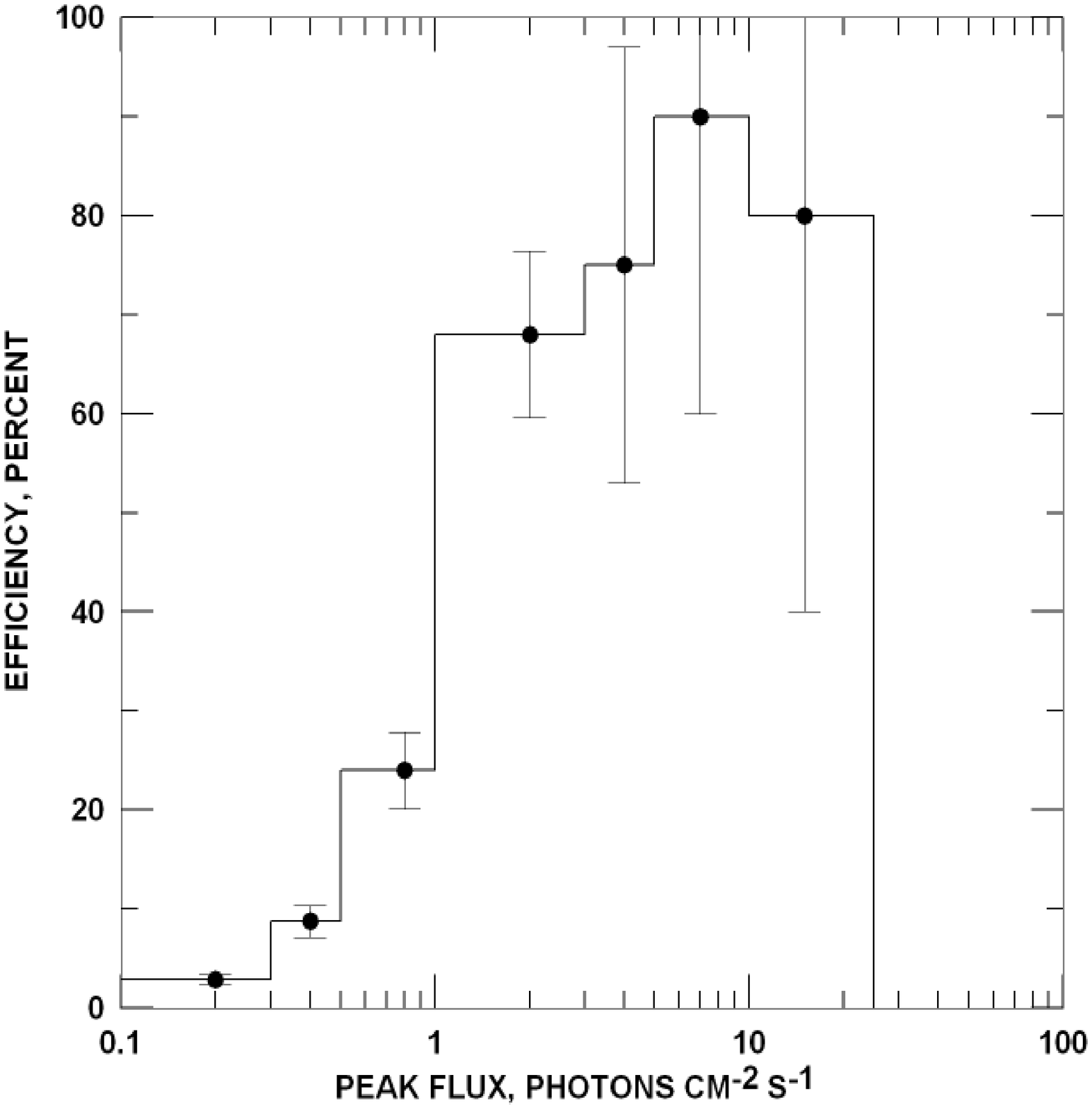}

\clearpage

\plotone{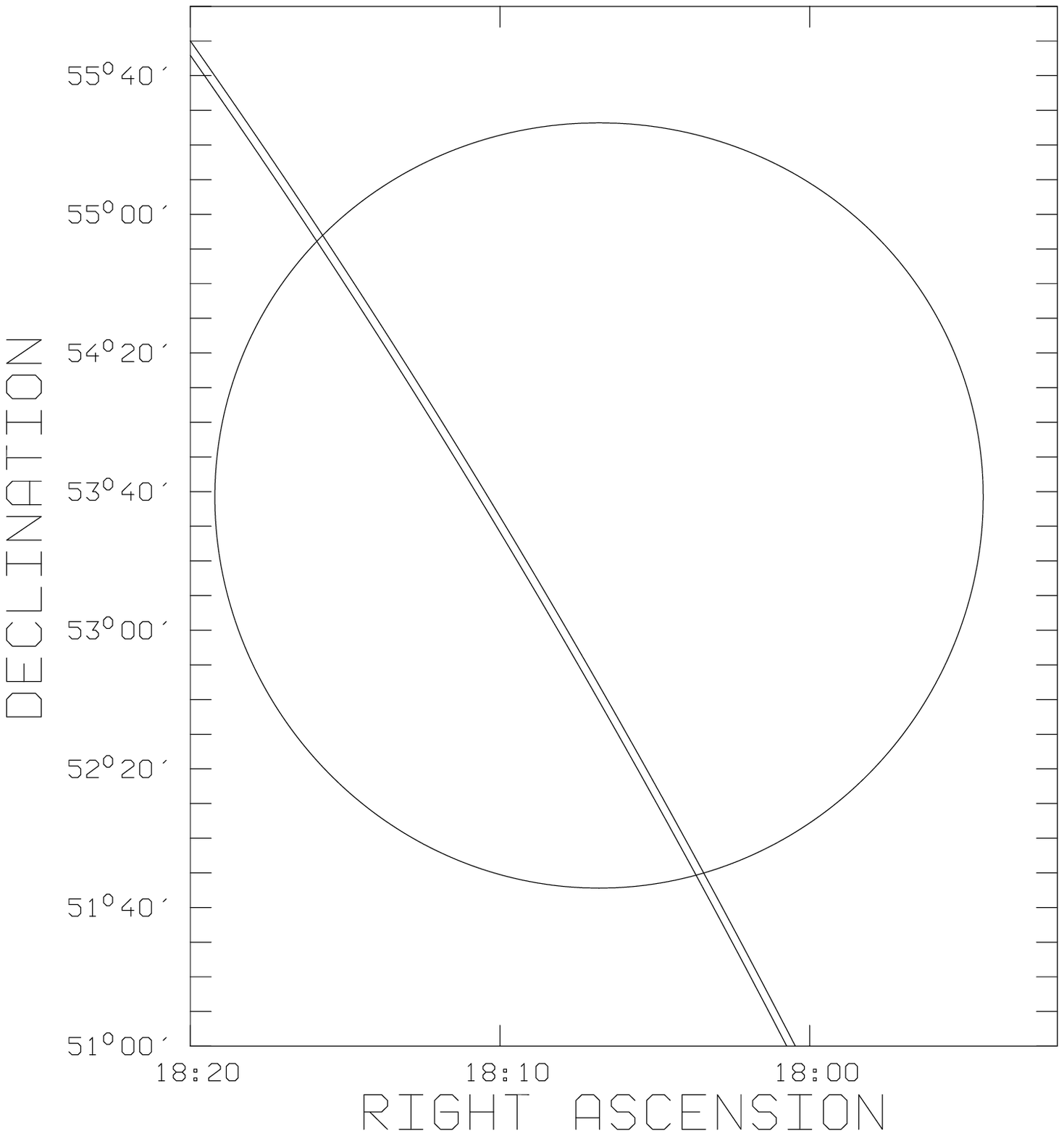}

\clearpage

\plotone{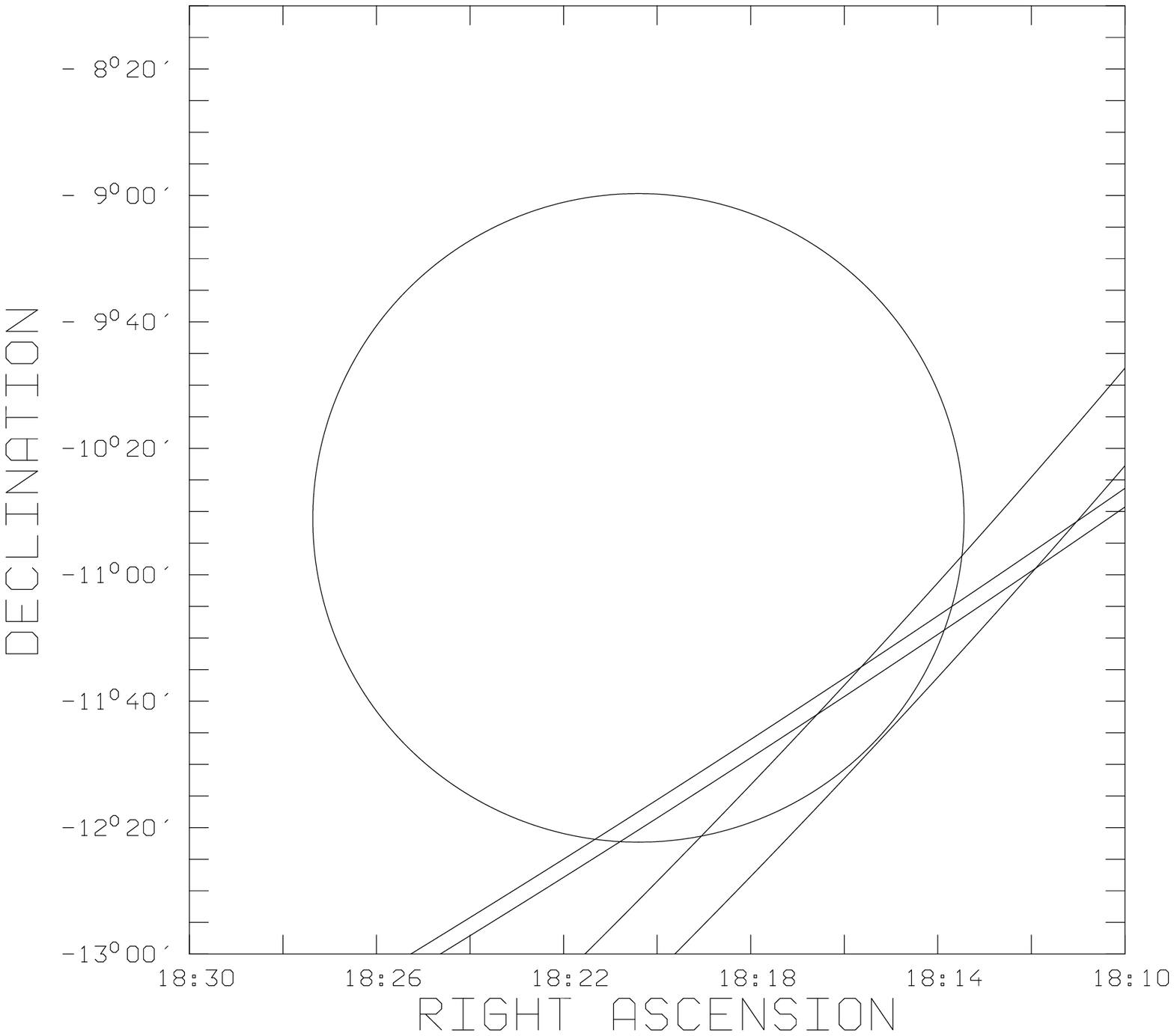}

\clearpage

\begin{deluxetable}{ccc}

\tablecaption{BATSE untriggered bursts confirmed by the IPN}

\tablehead{

\colhead{Date} & \colhead{UT} & \colhead{IPN spacecraft}

}

\startdata

 910601 & 62220 &  \it Ulysses \rm \\

 910830 & 05148 &  \it Ulysses \rm \\

 910908 & 33924 &  \it Ulysses \rm \\

 910910 & 14747 &  \it Ulysses \rm \\

 911029 & 17453 &  \it Ulysses, \rm PVO\tablenotemark{a}, \it Ginga\tablenotemark{b} \rm \\

 911120 & 43957 &  \it Ulysses \rm \\

 920109 & 23306 &  \it Ulysses \rm \\

 920216 & 58688 &  \it Ulysses \rm \\

 920303 & 24523 &  \it Ulysses \rm \\

 920622 & 23828 &  \it Ulysses \rm \\

 920626 & 64276 &  \it Ulysses \rm \\

 920717 & 57852 &  \it Ulysses \rm \\

 920903\tablenotemark{c}& 05728 &  \it Ulysses, \rm WATCH/GRANAT\tablenotemark{d} \\

 930118 & 64426 &  \it Ulysses \rm \\

 930209\tablenotemark{c} & 15737 &  \it Ulysses, \rm PHEBUS\tablenotemark{e} \\

 930408 & 06847 &  PHEBUS\tablenotemark{e} \\

 930424 & 38888 &  \it Ulysses \rm \\

 930506 & 55245 &  \it Ulysses \rm \\

 930626 & 07023 &  \it Ulysses \rm \\

 930710 & 13810 &  \it Ulysses \rm \\

 930909 & 45100 &  \it Ulysses \rm \\

 940213 & 07260 &  \it Ulysses \rm \\

 940222 & 08083 &  \it Ulysses \rm \\

 940311 & 44967 &  \it Ulysses \rm \\

 940710\tablenotemark{c} & 35477 &  \it Ulysses \rm \\

 940712 & 00070 &  \it Ulysses \rm \\

 940727 & 40865 &  \it Ulysses \rm \\

 940730 & 39690 &  \it Ulysses, \rm DMSP\tablenotemark{f}, SROSS-C \\

 940930 & 23017 &  \it Ulysses \rm \\

 941104 & 35178 &  \it Ulysses \rm \\

 950104 & 32438 & Konus-\it Wind \rm \\

 950111 & 46528 &  \it Ulysses \rm \\

 950131 & 78592 &  \it Ulysses, \rm  Konus-\it Wind \rm \\

 950203 & 08456 &  \it Ulysses, \rm SROSS-C \\

 950207 & 72568 & Konus-\it Wind \rm \\

 950211 & 15919 &  \it Ulysses, \rm  Konus-\it Wind \rm \\

 950224 & 33800 & Konus-\it Wind \rm \\

 950603 & 21257 & Konus-\it Wind \rm \\

 950611 & 21122 &  \it Ulysses \rm \\

 950614 & 00779 & Konus-\it Wind \rm \\

 950615 & 12104 &  \it Ulysses \rm \\

 950622 & 71470 &  \it Ulysses, \rm  Konus-\it Wind \rm \\

 950625 & 09685 & Konus-\it Wind \rm \\

 950722 & 64127 &  \it Ulysses, \rm  Konus-\it Wind \rm \\

 950723 & 73608 & Konus-\it Wind \rm \\

 950728 & 45743 & Konus-\it Wind \rm \\

 950730\tablenotemark{c}& 76147 &  \it Ulysses, \rm  Konus-\it Wind \rm \\

 950904\tablenotemark{c}& 52777 &  \it Ulysses \rm \\

 951001 & 41868 &  \it Ulysses, \rm  Konus-\it Wind \rm \\

 951005 & 14826 & Konus-\it Wind \rm \\

 951013 & 57096 & Konus-\it Wind \rm \\

 951112 & 67850 &  \it Ulysses \rm \\

 951124 & 25132 &  \it Ulysses, \rm  Konus-\it Wind \rm \\

 951213 & 32675 & Konus-\it Wind \rm \\

 951215 & 73379 & Konus-\it Wind \rm \\

 951218 & 28745 &  \it Ulysses, \rm  Konus-\it Wind \rm \\

 951231 & 77068 & Konus-\it Wind \rm \\

 960107 & 68607 & Konus-\it Wind \rm \\

 960115 & 31956 &  \it Ulysses, \rm  Konus-\it Wind \rm \\

 960123 & 43643 & Konus-\it Wind \rm \\

 960201 & 82195 &  \it Ulysses \rm \\

 960202 & 05968 &  \it Ulysses, \rm  Konus-\it Wind \rm \\

 960207 & 65033 &  \it Ulysses, \rm  Konus-\it Wind \rm \\

 960304 & 48776 &  \it Ulysses \rm \\

 960321 & 19663 & Konus-\it Wind \rm \\

 960418 & 08267 & Konus-\it Wind \rm \\

 960504 & 18779 & Konus-\it Wind \rm \\

 960602 & 42667 & Konus-\it Wind \rm \\

 960603 & 60930 & Konus-\it Wind \rm \\

 960614 & 83621 & Konus-\it Wind \rm \\

 960715 & 58326 & Konus-\it Wind \rm \\

 960725 & 63535 & \it BeppoSAX\tablenotemark{g} \rm \\

 960817 & 24647 & Konus-\it Wind \rm \\

 960826 & 58072 & Konus-\it Wind \rm \\

 960905 & 02568 & Konus-\it Wind \rm \\

 961017 & 23648 &  \it Ulysses \rm \\

 961023 & 07747 & \it BeppoSAX\tablenotemark{g} \rm \\

 961106 & 43031 & \it BeppoSAX\tablenotemark{g} \rm \\

 961107 & 12691 & Konus-\it Wind \rm \\

 961110 & 26976 & Konus-\it Wind, \rm  \it BeppoSAX\tablenotemark{g} \rm \\

 961113 & 80523 & Konus-\it Wind \rm \\

 961119\tablenotemark{c} & 21322 &  \it Ulysses, \rm  Konus-\it Wind \rm \\

 961119\tablenotemark{c} & 21536 & Konus-A \\

 961119\tablenotemark{c} & 26961 & Konus-A \\

 961120 & 30433 & \it BeppoSAX\tablenotemark{g} \rm \\

 961123 & 59316 & Konus-\it Wind \rm \\

 961208 & 68232 & \it BeppoSAX\tablenotemark{g} \rm \\

 961209 & 74677 &  \it Ulysses \rm \\

 961213 & 49966 &  \it Ulysses, \rm  Konus-\it Wind \rm \\

 961222 & 43207 & \it BeppoSAX\tablenotemark{g} \rm \\

 961224 & 36648 & Konus-\it Wind, \rm  \it BeppoSAX\tablenotemark{g} \rm \\

 970116 & 58238 &  \it Ulysses, \rm  Konus-\it Wind \rm \\

 970119 & 42607 & Konus-\it Wind \rm \\

 970221 & 13750 & \it BeppoSAX\tablenotemark{g} \rm \\

 970223 & 64885 & \it BeppoSAX\tablenotemark{g} \rm \\

 970311 & 30254 & \it BeppoSAX\tablenotemark{g} \rm \\

 970406 & 25471 &  \it Ulysses, \rm  Konus-\it Wind, \rm  \it BeppoSAX\tablenotemark{g} \rm \\

 970525 & 31783 & \it BeppoSAX\tablenotemark{g} \rm \\

 970610 & 36151 & \it BeppoSAX\tablenotemark{g} \rm \\

 970617 & 61459 & \it BeppoSAX\tablenotemark{g} \rm \\

 970720 & 68515 & Konus-\it Wind \rm \\

 970801 & 29048 &  \it Ulysses, \rm  Konus-\it Wind \rm \\

 970817 & 69692 & \it BeppoSAX\tablenotemark{g} \rm \\

 970825 & 40632 & \it BeppoSAX\tablenotemark{g} \rm \\

 970827 & 25872 & \it BeppoSAX\tablenotemark{g} \rm \\

 970926 & 79655 & Konus-\it Wind \rm \\

 971015 & 20356 & Konus-\it Wind \rm \\

 971017 & 01897 & Konus-\it Wind \rm \\

 971019 & 57427 & Konus-\it Wind \rm \\

 971027 & 09808 &  \it Ulysses, BeppoSAX\tablenotemark{g} \rm \\

 971028 & 75126 & \it BeppoSAX\tablenotemark{g} \rm \\

 971101 & 23483 &  \it Ulysses, \rm  Konus-\it Wind \rm \\

 971102 & 05581 & \it BeppoSAX\tablenotemark{g} \rm \\

 971103 & 27090 & \it BeppoSAX\tablenotemark{g} \rm \\

 971121 & 43992 &  \it Ulysses, \rm  Konus-\it Wind \rm \\

 971207 & 67900 &  \it Ulysses, \rm  Konus-\it Wind \rm \\

 971207 & 75492 &  \it Ulysses, \rm\it BeppoSAX\tablenotemark{g} \rm \\

 971228 & 53605 & \it BeppoSAX\tablenotemark{g} \rm \\

 971228 & 79012 & Konus-\it Wind, \rm NEAR \\

 980106 & 44231 & Konus-\it Wind \rm \\

 980205 & 19783 &  \it Ulysses, \rm  Konus-\it Wind, \rm  \it BeppoSAX\tablenotemark{g}, \rm  NEAR \\

 980207 & 58212 & Konus-\it Wind \rm \\

 980223 & 76640 & \it BeppoSAX\tablenotemark{g} \rm \\

 980226 & 41332 & \it BeppoSAX\tablenotemark{g} \rm \\

 980304 & 52863 & Konus-\it Wind, \rm  \it BeppoSAX\tablenotemark{g} \rm \\

 980329 & 55486 & \it BeppoSAX\tablenotemark{g} \rm \\

 980429 & 20493 & Konus-\it Wind \rm \\

 980518 & 67488 & \it BeppoSAX\tablenotemark{g} \rm \\

 980520 & 52002 & \it BeppoSAX\tablenotemark{g} \rm \\

 980523 & 31208 &  \it Ulysses, \rm  Konus-\it Wind \rm \\

 980602 & 46528 & Konus-\it Wind \rm \\

 980605 & 51131 & Konus-\it Wind, \rm  \it BeppoSAX\tablenotemark{g} \rm \\

 980613 & 17465 & \it BeppoSAX\tablenotemark{h} \rm \\

 980622\tablenotemark{c} & 51085 &  \it Ulysses \rm \\

 980626 & 70184 & Konus-\it Wind \rm \\

 980629 & 32377 &  \it Ulysses, \rm  Konus-\it Wind \rm \\

 980705 & 23165 & \it BeppoSAX\tablenotemark{g} \rm \\

 980706 & 63987 & Konus-\it Wind, \rm  \it BeppoSAX\tablenotemark{g} \rm \\

 980709 & 16963 & \it BeppoSAX\tablenotemark{g} \rm \\

 980712 & 18577 & \it BeppoSAX\tablenotemark{g} \rm \\

 980713 & 13301 & Konus-\it Wind, \rm  \it BeppoSAX\tablenotemark{g} \rm \\

 980715 & 35282 & \it BeppoSAX\tablenotemark{g} \rm \\

 980728 & 53879 & Konus-\it Wind \rm \\

 980728 & 55355 & Konus-\it Wind \rm \\

 980808 & 78791 & \it BeppoSAX\tablenotemark{g} \rm \\

 980810 & 15944 & \it BeppoSAX\tablenotemark{g} \rm \\

 980812 & 17640 & \it BeppoSAX\tablenotemark{g} \rm \\

 980812 & 18950 &  \it Ulysses, \rm  Konus-\it Wind, \rm  \it BeppoSAX\tablenotemark{g} \rm \\

 980907 & 40388 & \it BeppoSAX\tablenotemark{g} \rm \\

 980908 & 02480 & \it BeppoSAX\tablenotemark{g} \rm \\

 980913\tablenotemark{c} & 19983 &  \it Ulysses, \rm NEAR \\

 980916 & 73322 & \it BeppoSAX\tablenotemark{g} \rm \\

 980917 & 35279 & \it BeppoSAX\tablenotemark{g} \rm \\

 980923 & 30178 & \it BeppoSAX\tablenotemark{g} \rm \\

 981002 & 05466 & \it BeppoSAX\tablenotemark{g} \rm \\

 981018 & 01612 & \it BeppoSAX\tablenotemark{g} \rm \\

 981019 & 79603 &  \it Ulysses, \rm  Konus-\it Wind, \rm  \it BeppoSAX\tablenotemark{g} \rm \\

 981022 & 21682 & \it BeppoSAX\tablenotemark{g} \rm \\

 981022\tablenotemark{c}& 56447 &  \it Ulysses, \rm  Konus-\it Wind \rm \\

 981101 & 26940 &  \it Ulysses, \rm  Konus-\it Wind, \rm NEAR \\

 981106 & 38479 & \it BeppoSAX\tablenotemark{g} \rm \\

 981215 & 80709 & Konus-\it Wind, \rm NEAR \\

 981216 & 19755 & \it BeppoSAX\tablenotemark{g} \rm \\

 990104 & 39597 & \it BeppoSAX\tablenotemark{g} \rm \\

 990109 & 41054 &  \it Ulysses, \rm  Konus-\it Wind \rm \\

 990110\tablenotemark{c}& 31141 &  \it Ulysses \rm \\

 990128 & 37252 &  \it Ulysses, \rm  Konus-\it Wind, \rm  \it BeppoSAX\tablenotemark{g} \rm \\

 990204 & 30169 &  \it Ulysses, \rm  Konus-\it Wind \rm \\

 990305 & 34451 & Konus-\it Wind \rm \\

 990421 & 65775 &  \it Ulysses \rm \\

 990504 & 40929 & \it BeppoSAX\tablenotemark{g} \rm \\

 990509 & 74345 &  \it Ulysses \rm \\

 990526 & 47273 & \it BeppoSAX\tablenotemark{g} \rm \\

 990603 & 66686 & \it BeppoSAX\tablenotemark{g} \rm \\

 990606 & 11124 & Konus-\it Wind \rm \\

 990618 & 37636 & \it BeppoSAX\tablenotemark{g} \rm \\

 990621 & 43943 & \it BeppoSAX\tablenotemark{g} \rm \\

 990705 & 57685 &  \it Ulysses, \rm  Konus-\it Wind, \rm  \it BeppoSAX\tablenotemark{g}, \rm  NEAR \\

 990707 & 54801 &  \it Ulysses, \rm  Konus-\it Wind, \rm  \it BeppoSAX\tablenotemark{g} \rm \\

 990711 & 49110 & \it BeppoSAX\tablenotemark{g} \rm \\

 990719 & 79380 & \it BeppoSAX\tablenotemark{g} \rm \\

 990720 & 00025 & \it BeppoSAX\tablenotemark{g} \rm \\

 990725 & 41016 & \it BeppoSAX\tablenotemark{g} \rm \\

 990727 & 48288 & \it BeppoSAX\tablenotemark{g} \rm \\

 990803 & 57565 & \it BeppoSAX\tablenotemark{g} \rm \\

 990806 & 60168 & Konus-\it Wind \rm \\

 990828 & 70019 &  \it Ulysses, \rm  Konus-\it Wind \rm \\

 990917 & 52494 & \it BeppoSAX\tablenotemark{g} \rm \\

 990917 & 71095 & \it BeppoSAX\tablenotemark{g} \rm \\

 990919 & 49338 & Konus-\it Wind, \rm  \it BeppoSAX\tablenotemark{g}, \rm  NEAR \\

 990919 & 86038 & Konus-\it Wind \rm \\

 990926 & 32653 & NEAR \\

 991002 & 15031 & \it BeppoSAX\tablenotemark{g} \rm \\

 991004 & 22825 & \it BeppoSAX\tablenotemark{g} \rm \\

 991005 & 15265 &  \it Ulysses \rm \\

 991011 & 35968 & Konus-\it Wind, \rm  \it BeppoSAX\tablenotemark{g} \rm \\

 991120 & 27069 & NEAR \\

 991205 & 82651 & \it BeppoSAX\tablenotemark{g} \rm \\

 991217 & 21782 & \it BeppoSAX\tablenotemark{g} \rm \\

 000102 & 27709 &  \it Ulysses \rm \\

 000205 & 45486 &  \it Ulysses, \rm\it BeppoSAX\tablenotemark{g} \rm \\

 000206 & 09183 &  \it Ulysses, \rm\it BeppoSAX\tablenotemark{g} \rm \\

 000210 & 14030 & Konus-\it Wind \rm \\

 000211 & 45217 & \it BeppoSAX\tablenotemark{g} \rm \\

 000224 & 82209 & \it BeppoSAX\tablenotemark{g} \rm \\

 000318 & 12931 & Konus-\it Wind \rm \\

 000403 & 13199 &  \it Ulysses, \rm  Konus-\it Wind, \rm NEAR \\

 000405 & 77386 &  \it Ulysses \rm \\

 000420 & 61374 &  \it Ulysses, \rm  Konus-\it Wind \rm \\

 000502 & 54060 & \it BeppoSAX\tablenotemark{g} \rm \\

 000511 & 66298 &  \it Ulysses, \rm  Konus-\it Wind \rm \\

\enddata

\tablenotetext{a}{J. Laros, private communication, 1991}

\tablenotetext{b}{T. Murakami, private communication, 1991}

\tablenotetext{c}{See section 5}

\tablenotetext{d}{Sazonov et al. 1998}

\tablenotetext{e}{Tkachenko et al. 1998}

\tablenotetext{f}{J. Terrell, private communication, 1995}

\tablenotetext{g}{Guidorzi et al. 2004}

\tablenotetext{h}{Piro \& Costa 1998}

\end{deluxetable}

\clearpage

\begin{deluxetable}{cccccc}




\tablecaption{\it IPN annuli}

\tablehead{

\colhead{Date}&\colhead{UT}&\colhead{$\alpha_{2000, IPN}$}&

\colhead{$\delta_{2000, IPN}$}&\colhead{R$_{IPN1}$}&\colhead{$\delta R_{IPN}$}

}

\startdata

910601 & 62220 & 307.008 & -20.544 & 31.555 &  0.602 \\

910830 & 05148 & 152.409 &  12.566 & 48.690 &  0.306 \\

910908 & 33924 & 154.638 &  11.730 & 84.755 &  0.092 \\

910910 & 14747 & 155.049 &  11.574 & 19.634 &  0.072 \\

911029 & 17453 & 344.478 &  -7.880 & 29.352 &  0.050 \\

911120 & 43957 & 167.141 &   6.825 & 86.994 &  0.049 \\

920109 & 23306 & 347.750 &  -6.711 & 47.721 &  0.173 \\

920216 & 58688 & 342.389 &  -8.336 & 32.244 &  0.135 \\

920303 & 24523 & 338.881 &  -8.709 & 21.076 &  0.083 \\

920622 & 23828 & 150.168 &   6.668 & 48.411 &  0.135 \\

920626 & 64276 & 330.525 &  -6.378 & 65.379 &  0.045 \\

920717 & 57852 & 152.612 &   4.858 & 77.394 &  0.050 \\

920903& 05728 & 338.736 &  -0.545 & 54.228 &  0.063 \\

930118 & 64426 & 163.857 & -13.874 & 54.969 &  0.035 \\

930209 & 15737 & 159.547 & -14.829 & 70.983 &  0.050 \\

930424 & 38888 & 144.212 & -12.262 & 78.603 &  0.549 \\

930506 & 55245 & 323.364 &  11.717 & 66.872 &  0.992 \\

930626 & 07023 & 325.198 &  11.491 & 67.667 &  0.374 \\

930710 & 13810 & 326.814 &  12.090 & 82.054 &  0.024 \\

930909 & 45100 & 336.355 &  17.978 & 87.746 &  0.038 \\

940213 & 07260 & 151.385 & -51.418 & 39.877 &  0.189 \\

940222 & 08083 & 146.406 & -52.260 & 86.434 &  0.118 \\

940311 & 44967 & 136.417 & -52.192 & 85.580 &  0.439 \\

940710 & 35477 & 128.502 & -39.575 & 25.746 &  0.750 \\

940712 & 00070 & 128.902 & -39.677 & 44.911 &  0.058 \\

940727 & 40865 & 313.160 &  41.197 & 47.661 &  0.206 \\

940730 & 39690 & 134.062 & -41.599 & 35.990 &  0.072 \\

940930 & 23017 & 341.171 &  58.806 & 65.096 &  0.510 \\

941104 & 35178 &  26.375 &  73.600 & 57.376 &  0.187 \\

950111 & 46528 & 322.648 & -44.901 & 66.079 &  0.119 \\

950131 & 78592 & 331.984 & -31.567 & 29.273 &  0.527 \\

950203 & 08456 & 332.839 & -30.167 & 89.992 &  0.030 \\

950211 & 15919 & 335.834 & -25.077 & 30.200 &  0.130 \\

950611 & 21122 & 196.445 & -57.824 & 26.966 &  1.033 \\

950615 & 12104 & 198.715 & -60.874 & 56.631 &  0.518 \\

950622 & 71470 & 204.112 & -66.756 & 68.801 &  0.141 \\

950722 & 64127 &  98.551 &  83.396 & 57.905 &  0.379 \\

950730& 76147 & 320.422 & -82.230 & 77.858 &  0.039 \\

950904 & 52777 & 190.636 &  67.767 & 36.463 &  0.144 \\

951001 & 41868 & 202.666 &  60.024 & 80.559 &  0.068 \\

951112 & 67850 & 216.222 &  54.930 & 39.289 &  0.202 \\

951124 & 25132 & 219.201 &  55.029 & 38.194 &  1.044 \\

951218 & 28745 &  44.049 & -57.317 & 39.343 &  0.089 \\

960115 & 31956 &  45.389 & -63.277 & 45.746 &  0.090 \\

960201 & 82195 &  41.130 & -68.152 & 88.655 &  0.146 \\

960202 & 05968 & 221.086 &  68.185 & 49.051 &  0.164 \\

960207 & 65033 & 218.206 &  69.749 & 73.961 &  0.089 \\

960304 & 48776 & 192.220 &  74.434 & 75.937 &  0.404 \\

961017 & 23648 & 174.945 &  32.043 & 15.768 &  0.302 \\

961209 & 74677 & 179.842 &  31.294 &  2.542 &  5.564 \\

961213 & 49966 & 179.845 &  31.446 & 24.486 &  0.129 \\

970116 & 58238 & 176.897 &  33.808 & 84.760 &  0.143 \\

970406 & 25471 & 335.943 & -35.323 & 42.658 &  0.277 \\

970801 & 29048 & 337.672 & -22.111 & 38.624 &  0.222 \\

971027 & 09808 & 169.177 &  13.952 & 58.403 &  0.013 \\

971101 & 23483 & 349.656 & -13.606 & 27.793 &  0.014 \\

971121 & 43992 & 351.033 & -12.487 & 73.989 &  0.115 \\

971207 & 67900 & 171.412 &  11.889 & 75.290 &  0.226 \\

971207 & 75492 & 171.412 &  11.887 & 62.197 &  0.051 \\

971228 & 79012 &  83.395 &  20.899 & 81.666 &  2.305 \\

980205 & 19783 & 165.363 &  12.132 & 53.914 &  0.040 \\

       &       & 183.317 & -72.020 & 38.790 &  2.212 \\

980523 & 31208 & 329.478 & -11.682 & 36.270 &  0.183 \\

980622 & 51085 & 330.752 &  -9.840 & 77.221 &  0.029 \\

980629 & 32377 & 331.305 &  -9.351 & 80.020 &  0.019 \\

980812 & 18950 & 336.276 &  -5.670 & 38.173 &  0.061 \\

980913 & 19983 & 340.559 &  -2.546 & 31.980 &  0.278 \\

       &       &  67.928 &  25.235 & 56.955 &  0.483 \\

981019 & 79603 & 344.985 &   1.273 & 45.039 &  0.029 \\

981022 & 56447 & 345.260 &   1.560 & 58.633 &  0.012 \\

981101 & 26940 & 346.143 &   2.571 & 61.551 &  0.180 \\

       &       & 275.722 & -23.305 & 44.973 &  0.085 \\

981215 & 80709 & 301.383 & -18.295 & 16.297 &  0.223 \\

990109 & 41054 & 345.757 &   8.751 & 60.734 &  0.016 \\

990110 & 31141 & 345.649 &   8.805 & 58.037 &  0.004 \\

990128 & 37252 & 342.818 &   9.649 & 62.261 &  0.007 \\

990204 & 30169 & 161.450 &  -9.832 & 73.846 &  0.169 \\

990421 & 65775 & 146.070 &  -7.792 & 55.096 &  0.026 \\

990509 & 74345 & 144.820 &  -7.244 & 81.947 &  0.375 \\

990705 & 57685 & 147.515 &  -8.031 & 76.306 &  0.004 \\

       &       & 167.925 & -19.482 & 71.632 &  0.008 \\

990707 & 54801 & 147.737 &  -8.132 & 56.409 &  0.010 \\

990828 & 70019 & 155.115 & -12.504 & 67.522 &  0.024 \\

990919 & 49338 & 149.773 &  12.594 & 74.222 &  0.082 \\

990926 & 32653 & 335.529 &  -9.725 & 14.580 &  1.019 \\

991005 & 15265 & 340.866 &  17.520 & 18.198 &  0.110 \\

991120 & 27069 & 198.957 & -13.840 & 45.024 &  0.095 \\

000102 & 27709 & 165.147 & -34.791 & 51.204 &  0.236 \\

000205 & 45486 & 156.800 & -40.955 & 70.209 &  0.128 \\

000206 & 09183 & 336.592 &  41.031 & 51.005 &  0.395 \\

000403 & 13199 & 314.499 &  40.072 & 63.855 &  0.040 \\

       &       & 308.246 &  19.750 & 46.194 &  0.167 \\

000405 & 77386 & 133.665 & -39.692 & 57.653 &  0.304 \\

000420 & 61374 & 310.666 &  37.563 & 86.354 &  0.181 \\

000511 & 66298 & 309.117 &  34.756 & 75.388 &  0.712 \\

\enddata

\end{deluxetable}

\clearpage

\noindent

\clearpage

\begin{deluxetable}{cccccc}

\tabletypesize{\scriptsize}

\tablecaption{\it IPN error boxes}

\tablewidth{0pt}

\tablehead{

\colhead{Date}&\colhead{UT}&\colhead{$\alpha_{2000}$}&

\colhead{$\delta_{2000}$}&\colhead{$\sigma_{sys+stat, B}$}

}

\startdata

910601 & 62220 & 297.900 &   8.300 &  2.330 \\

       &       & 296.058 &   9.756 &        \\

       &       & 298.589 &  10.529 &        \\

       &       & 295.545 &   8.284 &        \\

       &       & 299.858 &   9.600 &        \\

910830 &  5148 & 202.100 &  14.700 &  2.130 \\

       &       & 202.699 &  12.651 &        \\

       &       & 202.991 &  16.650 &        \\

       &       & 202.060 &  12.570 &        \\

       &       & 202.358 &  16.815 &        \\

910908 & 33924 & 230.800 & -32.000 &  2.260 \\

       &       & 230.053 & -34.172 &        \\

       &       & 231.638 & -29.857 &        \\

       &       & 229.842 & -34.113 &        \\

       &       & 231.434 & -29.806 &        \\

910910 & 14747 & 150.700 &  30.300 &  2.000 \\

       &       & 148.384 &  30.285 &        \\

       &       & 152.795 &  31.170 &        \\

       &       & 148.393 &  30.134 &        \\

       &       & 152.864 &  31.031 &        \\

911029 & 17453 &  11.300 & -18.700 &  2.130 \\

       &       &  11.838 & -20.769 &        \\

       &       &  13.048 & -17.368 &        \\

       &       &  11.712 & -20.794 &        \\

       &       &  12.966 & -17.277 &        \\

911120 & 43957 &  74.300 &  35.100 &  3.050 \\

       &       &  76.296 &  32.540 &        \\

       &       &  75.564 &  37.976 &        \\

       &       &  76.405 &  32.601 &        \\

       &       &  75.694 &  37.937 &        \\

920109 & 23306 & 345.500 &  33.800 & 11.410 \\

       &       & 332.965 &  39.182 &        \\

       &       & 357.208 &  40.384 &        \\

       &       & 332.726 &  38.745 &        \\

       &       & 357.521 &  39.973 &        \\

920216 & 58688 & 322.900 &  18.200 &  2.260 \\

       &       & 321.367 &  16.478 &        \\

       &       & 325.055 &  19.170 &        \\

       &       & 321.597 &  16.313 &        \\

       &       & 325.162 &  18.913 &        \\

920303 & 24523 & 318.400 & -10.900 &  4.220 \\

       &       & 318.255 & -15.118 &        \\

       &       & 317.616 &  -6.752 &        \\

       &       & 318.434 & -15.120 &        \\

       &       & 317.788 &  -6.724 &        \\

920622 & 23828 & 147.100 & -39.400 &  2.720 \\

       &       & 148.636 & -41.858 &        \\

       &       & 145.100 & -41.656 &        \\

       &       & 149.203 & -41.600 &        \\

       &       & 144.594 & -41.337 &        \\

920626 & 64276 &   1.000 &  50.900 &  2.060 \\

       &       &   0.000 &   0.000 &        \\

       &       &   0.000 &   0.000 &        \\

       &       &   0.000 &   0.000 &        \\

       &       &   0.000 &   0.000 &        \\

920717 & 57852 & 159.300 & -71.600 &  2.000 \\

       &       & 153.631 & -72.582 &        \\

       &       & 165.497 & -72.133 &        \\

       &       & 153.477 & -72.483 &        \\

       &       & 165.564 & -72.026 &        \\

920903 & 05728 & 299.100 &  28.800 &  1.890 \\

       &       & 295.016 &  35.387 &        \\

       &       & 295.583 &  36.120 &        \\

       &       & 295.067 &  35.214 &        \\

       &       & 295.799 &  36.158 &        \\

930118 & 64426 & 219.200 & -32.600 &  2.560 \\

       &       & 220.633 & -34.866 &        \\

       &       & 221.151 & -30.652 &        \\

       &       & 220.541 & -34.904 &        \\

       &       & 221.075 & -30.599 &        \\

930209 & 15737 & 239.300 & -58.200 &  1.750 \\

       &       & 237.240 & -59.590 &        \\

       &       & 237.460 & -56.756 &        \\

       &       & 237.051 & -59.508 &        \\

       &       & 237.272 & -56.830 &        \\

930424 & 38888 & 230.100 & -57.600 & 2.260  \\

       &       & 233.507 & -58.979 &        \\

       &       & 232.932 & -55.956 &        \\

       &       & 231.473 & -59.744 &        \\

       &       & 230.887 & -55.382 &        \\

930506 & 55245 & 259.300 &  34.400 & 11.710 \\

       &       & 252.549 &  24.267 &        \\

       &       & 253.061 &  45.092 &        \\

       &       & 254.773 &  23.377 &        \\

       &       & 256.039 &  45.844 &        \\

930626 & 07023 &  21.800 & -40.200 & 10.320 \\

       &       &   8.784 & -43.846 &        \\

       &       &  21.353 & -29.886 &        \\

       &       &   8.494 & -42.978 &        \\

       &       &  20.326 & -29.950 &        \\

930710 & 13810 & 316.500 & -69.200 &  2.060 \\

       &       & 322.028 & -69.921 &        \\

       &       & 310.697 & -69.198 &        \\

       &       & 322.071 & -69.873 &        \\

       &       & 310.705 & -69.149 &        \\

930909 & 45100 & 244.000 &   9.400 &  2.000 \\

       &       & 245.880 &   8.658 &        \\

       &       & 245.084 &  11.092 &        \\

       &       & 245.923 &   8.773 &        \\

       &       & 245.187 &  11.023 &        \\

940213 & 07260 & 191.900 & -28.100 &  2.130 \\

       &       & 194.111 & -27.262 &        \\

       &       & 192.498 & -26.038 &        \\

       &       & 194.303 & -27.915 &        \\

       &       & 191.751 & -25.974 &        \\

940222 &  8083 & 200.200 &  22.500 &  4.400 \\

       &       & 203.083 &  19.022 &        \\

       &       & 195.460 &  23.007 &        \\

       &       & 202.862 &  18.872 &        \\

       &       & 195.441 &  22.750 &        \\

940311 & 44967 & 105.200 &  26.400 &  2.130 \\

       &       &   0.000 &   0.000 &        \\

       &       &   0.000 &   0.000 &        \\

       &       & 103.541 &  27.936 &        \\

       &       & 105.460 &  28.518 &        \\

940710 & 35477 &  99.400 & -33.300 &  3.760 \\

       &       &  95.439 & -35.148 &        \\

       &       &  98.463 & -29.626 &        \\

       &       &  96.866 & -36.433 &        \\

       &       & 100.452 & -29.648 &        \\

940712 & 00070 &  66.700 & -73.100 &  2.800 \\

       &       &  66.982 & -75.899 &        \\

       &       &  63.191 & -70.522 &        \\

       &       &  67.472 & -75.892 &        \\

       &       &  63.527 & -70.480 &        \\

940727 & 40865 & 312.500 &  -1.900 &  2.260 \\

       &       &   0.000 &   0.000 &        \\

       &       &   0.000 &   0.000 &        \\

       &       &   0.000 &   0.000 &        \\

       &       &   0.000 &   0.000 &        \\

940730 & 39690 &  79.700 & -61.400 &  1.750 \\

       &       &  83.263 & -61.843 &        \\

       &       &  82.831 & -60.531 &        \\

       &       &   0.000 &   0.000 &        \\

       &       &   0.000 &   0.000 &        \\

940930 & 23017 &  78.700 &  32.000 &  3.760 \\

       &       &  74.499 &  30.863 &        \\

       &       &  81.554 &  34.911 &        \\

       &       &  74.267 &  31.968 &        \\

       &       &  80.379 &  35.492 &        \\

941104 & 35178 & 345.300 &  18.500 &  5.920 \\

       &       & 351.494 &  19.349 &        \\

       &       & 340.023 &  21.742 &        \\

       &       & 351.427 &  19.743 &        \\

       &       & 340.265 &  22.072 &        \\

950111 & 46528 &  24.300 &  -5.800 &  1.790 \\

       &       &  25.708 &  -6.916 &        \\

       &       &  23.085 &  -4.481 &        \\

       &       &  25.546 &  -7.092 &        \\

       &       &  22.919 &  -4.654 &        \\

950131 & 78592 & 319.400 & -58.800 & 6.790  \\

       &       & 332.034 & -61.367 &        \\

       &       & 307.773 & -56.126 &        \\

       &       & 332.489 & -60.312 &        \\

       &       & 308.789 & -55.212 &        \\

950203 & 08456 & 274.600 &  42.700 &  2.000 \\

       &       & 273.162 &  41.011 &        \\

       &       & 276.802 &  43.897 &        \\

       &       & 273.232 &  40.979 &        \\

       &       & 276.851 &  43.846 &        \\

950211 & 15919 & 325.600 &   3.300 &  1.750 \\

       &       & 323.883 &   2.949 &        \\

       &       & 327.178 &   4.064 &        \\

       &       & 323.955 &   2.698 &        \\

       &       & 327.274 &   3.822 &        \\

950611 & 21122 & 269.800 & -69.500 &  2.480 \\

       &       &   0.000 &   0.000 &        \\

       &       &   0.000 &   0.000 &        \\

       &       &   0.000 &   0.000 &        \\

       &       &   0.000 &   0.000 &        \\

950615 & 12104 & 310.200 & -48.500 &  6.110 \\

       &       & 317.991 & -52.050 &        \\

       &       & 302.386 & -45.503 &        \\

       &       & 316.943 & -52.875 &        \\

       &       & 301.692 & -46.428 &        \\

950622 & 71470 & 213.400 &  -1.500 &  4.680 \\

       &       & 216.922 &   1.582 &        \\

       &       & 210.342 &   2.043 &        \\

       &       & 217.169 &   1.275 &        \\

       &       & 210.055 &   1.773 &        \\

950722 & 64127 &  74.200 &  25.800 &  3.580 \\

       &       &  78.163 &  25.551 &        \\

       &       &  70.224 &  25.950 &        \\

       &       &  78.144 &  26.310 &        \\

       &       &  70.337 &  26.703 &        \\

950730 & 76147 & 218.300 & -13.400 &  2.193 \\

       &       & 216.499 & -14.090 &        \\

       &       & 220.702 & -13.533 &        \\

       &       & 216.511 & -14.168 &        \\

       &       & 220.711 & -13.611 &        \\

950904 & 52777 & 197.100 &  33.300 &  2.190 \\

       &       & 196.048 &  31.299 &        \\

       &       & 198.492 &  31.452 &        \\

       &       & 195.522 &  31.561 &        \\

       &       & 198.968 &  31.778 &        \\

951001 & 41868 & 127.100 &  -0.300 &  2.970 \\

       &       & 129.732 &   1.076 &        \\

       &       & 126.889 &   2.663 &        \\

       &       & 129.600 &   1.304 &        \\

       &       & 127.153 &   2.670 &        \\

951112 & 67850 & 225.700 &  16.900 &  1.940 \\

       &       & 223.970 &  15.896 &        \\

       &       & 227.667 &  16.437 &        \\

       &       & 223.782 &  16.281 &        \\

       &       & 227.727 &  16.858 &        \\

951124 & 25132 & 157.300 &  47.600 &  2.190 \\

       &       & 157.526 &  45.416 &        \\

       &       & 154.289 &  48.462 &        \\

       &       & 160.092 &  46.514 &        \\

       &       & 156.796 &  49.765 &        \\

951218 & 28745 &  20.700 & -20.700 &  2.000 \\

       &       &  19.312 & -22.227 &        \\

       &       &  22.801 & -21.085 &        \\

       &       &  19.491 & -22.354 &        \\

       &       &  22.748 & -21.288 &        \\

960115 & 31956 & 132.300 & -50.200 &  4.030 \\

       &       & 137.695 & -52.409 &        \\

       &       & 127.935 & -47.375 &        \\

       &       & 137.535 & -52.562 &        \\

       &       & 127.746 & -47.504 &        \\

960201 & 82195 & 170.700 & -15.600 &  6.400 \\

       &       & 177.126 & -17.328 &        \\

       &       & 164.395 & -13.665 &        \\

       &       & 177.042 & -17.609 &        \\

       &       & 164.306 & -13.944 &        \\

960202 & 05968 & 104.400 &  54.600 &  2.060 \\

       &       & 107.037 &  53.246 &        \\

       &       & 100.892 &  54.993 &        \\

       &       & 107.379 &  53.511 &        \\

       &       & 101.036 &  55.318 &        \\

960207 & 65033 &   9.800 &  35.800 &  2.260 \\

       &       &   8.257 &  33.928 &        \\

       &       &  12.029 &  34.464 &        \\

       &       &   8.022 &  34.072 &        \\

       &       &  12.193 &  34.665 &        \\

960304 & 48776 &  72.800 &  23.400 &  3.050 \\

       &       &  75.184 &  21.292 &        \\

       &       &  69.613 &  22.567 &        \\

       &       &  75.734 &  21.995 &        \\

       &       &  69.476 &  23.427 &        \\

961017 & 23648 & 162.800 &  42.400 &  4.030 \\

       &       & 157.992 &  40.590 &        \\

       &       & 164.912 &  46.136 &        \\

       &       & 158.495 &  40.001 &        \\

       &       & 165.805 &  45.805 &        \\

961209 & 74677 & 188.600 &  33.400 &  5.630 \\

       &       & 188.241 &  27.778 &        \\

       &       & 184.992 &  38.211 &        \\

       &       & 183.130 &  30.222 &        \\

       &       & 181.856 &  33.794 &        \\

961213 & 49966 & 186.300 &  57.300 &  2.720 \\

       &       & 189.274 &  55.140 &        \\

       &       & 181.933 &  56.019 &        \\

       &       & 188.793 &  54.961 &        \\

       &       & 182.254 &  55.744 &        \\

970116 & 58238 & 181.300 & -49.300 &  2.560 \\

       &       & 178.432 & -51.084 &        \\

       &       & 184.499 & -50.830 &        \\

       &       & 178.071 & -50.803 &        \\

       &       & 184.795 & -50.520 &        \\

970406 & 25471 & 287.300 & -28.200 & 4.120  \\

       &       & 284.199 & -31.319 &        \\

       &       & 287.704 & -24.096 &        \\

       &       & 284.742 & -31.673 &        \\

       &       & 288.335 & -24.186 &        \\

970801 & 29048 & 311.600 & -54.200 &  2.260 \\

       &       & 312.760 & -56.361 &        \\

       &       & 307.838 & -53.740 &        \\

       &       & 313.585 & -56.156 &        \\

       &       & 308.119 & -53.267 &        \\

971027 & 09808 & 194.600 &  67.500 &  1.630 \\

       &       &   0.000 &   0.000 &        \\

       &       &   0.000 &   0.000 &        \\

       &       &   0.000 &   0.000 &        \\

       &       &   0.000 &   0.000 &        \\

971101 & 23483 & 359.000 &  13.100 &  1.840 \\

       &       &   0.531 &  12.026 &        \\

       &       & 357.113 &  13.202 &        \\

       &       &   0.513 &  12.003 &        \\

       &       & 357.112 &  13.173 &        \\

971121 & 43992 & 288.000 & -71.000 &  3.310 \\

       &       &   0.000 &   0.000 &        \\

       &       &   0.000 &   0.000 &        \\

       &       &   0.000 &   0.000 &        \\

       &       &   0.000 &   0.000 &        \\

971207 & 67900 &  91.200 &  56.000 &  5.150 \\

       &       &  89.818 &  50.916 &        \\

       &       &  89.893 &  61.105 &        \\

       &       &  90.537 &  50.865 &        \\

       &       &  90.833 &  61.146 &        \\

971207 & 75492 & 131.400 &  71.100 &  5.540 \\

       &       & 125.571 &  65.980 &        \\

       &       & 148.573 &  72.644 &        \\

       &       & 125.820 &  65.945 &        \\

       &       & 148.619 &  72.537 &        \\

971228 & 79012 &  50.600 & -58.100 &  3.580 \\

       &       &  56.348 & -60.131 &        \\

       &       &  44.681 & -56.488 &        \\

       &       &   0.000 &   0.000 &        \\

       &       &   0.000 &   0.000 &        \\

980205 & 19783 & 146.400 & -37.400 &  2.449 \\

       &       & 141.408 & -36.889 &        \\

       &       & 149.002 & -39.630 &        \\

       &       & 141.537 & -36.855 &        \\

       &       & 149.169 & -39.591 &        \\

980523 & 31208 & 336.100 & -50.600 &  5.440 \\

       &       & 328.652 & -48.129 &        \\

       &       & 341.734 & -46.631 &        \\

       &       & 328.989 & -47.767 &        \\

       &       & 341.252 & -46.362 &        \\

980622 & 51085 & 247.000 & -49.900 &  5.350 \\

       &       & 248.967 & -55.115 &        \\

       &       & 249.002 & -44.724 &        \\

       &       & 249.068 & -55.100 &        \\

       &       & 249.083 & -44.739 &        \\

980629 & 32377 & 271.700 &  53.600 &  1.840 \\

       &       & 270.855 &  51.833 &        \\

       &       & 273.931 &  54.899 &        \\

       &       & 270.917 &  51.822 &        \\

       &       & 273.978 &  54.870 &        \\

980812 & 18950 & 337.300 &  33.900 &  3.140 \\

       &       & 340.558 &  32.346 &        \\

       &       & 333.941 &  32.499 &        \\

       &       & 340.474 &  32.231 &        \\

       &       & 334.018 &  32.381 &        \\

980913 & 19983 &  10.800 &   6.500 &  3.580 \\

       &       &  12.306 &   3.250 &        \\

       &       &  10.380 &  10.056 &        \\

       &       &  11.775 &   3.054 &        \\

       &       &   9.818 &   9.945 &        \\

981019 & 79603 & 318.700 & -36.700 &  3.220 \\

       &       & 322.038 & -38.539 &        \\

       &       & 315.772 & -34.531 &        \\

       &       & 322.079 & -38.490 &        \\

       &       & 315.821 & -34.489 &        \\

981022 & 56447 & 287.900 &   8.000 &  6.110 \\

       &       & 286.585 &   2.032 &        \\

       &       & 287.253 &  14.077 &        \\

       &       & 286.609 &   2.027 &        \\

       &       & 287.278 &  14.079 &        \\

981101 & 26940 & 290.700 &  18.400 &  2.260 \\

       &       & 285.480 &  20.739 &        \\

       &       & 285.453 &  20.570 &        \\

       &       & 285.856 &  20.658 &        \\

       &       & 285.829 &  20.489 &        \\

981215 & 80709 & 298.000 &  -2.800 &  2.720 \\

       &       & 295.278 &  -2.895 &        \\

       &       & 300.530 &  -1.796 &        \\

       &       & 295.333 &  -3.352 &        \\

       &       & 300.663 &  -2.236 &        \\

990109 & 41054 & 319.500 & -49.200 &  1.940 \\

       &       & 322.111 & -48.305 &        \\

       &       & 319.282 & -47.265 &        \\

       &       & 322.068 & -48.254 &        \\

       &       & 319.369 & -47.262 &        \\

990110 & 31141 & 287.100 &   9.700 &  3.223 \\

       &       & 287.081 &   6.477 &        \\

       &       & 286.582 &  12.883 &        \\

       &       & 287.088 &   6.477 &        \\

       &       & 286.589 &  12.884 &        \\

990128 & 37252 & 304.900 & -41.900 &  1.680 \\

       &       & 306.488 & -43.105 &        \\

       &       & 303.203 & -40.804 &        \\

       &       & 306.503 & -43.094 &        \\

       &       & 303.216 & -40.793 &        \\

990204 & 30169 & 131.600 &  63.200 &  3.490 \\

       &       & 130.012 &  59.793 &        \\

       &       & 138.763 &  62.040 &        \\

       &       & 131.050 &  59.720 &        \\

       &       & 138.262 &  61.569 &        \\

990421 & 65775 &  93.000 & -44.000 &  2.640 \\

       &       &   0.000 &   0.000 &        \\

       &       &   0.000 &   0.000 &        \\

       &       &   0.000 &   0.000 &        \\

       &       &   0.000 &   0.000 &        \\

990509 & 74345 & 234.000 & -25.300 &  7.280 \\

       &       & 230.251 & -31.794 &        \\

       &       & 229.206 & -19.522 &        \\

       &       & 229.333 & -31.312 &        \\

       &       & 228.459 & -20.114 &        \\

990705 & 57685 &  79.100 & -72.300 &  1.750 \\

       &       &  77.451 & -72.112 &        \\

       &       &  77.488 & -72.150 &        \\

       &       &  77.460 & -72.090 &        \\

       &       &  77.497 & -72.127 &        \\

990707 & 54801 & 102.700 & -53.000 &  1.840 \\

       &       & 105.730 & -53.289 &        \\

       &       & 103.292 & -51.196 &        \\

       &       & 105.736 & -53.261 &        \\

       &       & 103.336 & -51.202 &        \\

990828 & 70019 & 221.000 & -66.600 &  2.560 \\

       &       & 215.426 & -67.990 &        \\

       &       & 219.021 & -64.176 &        \\

       &       & 215.346 & -67.937 &        \\

       &       & 218.889 & -64.195 &        \\

990919 & 49338 &  69.400 &  74.000 &  2.720 \\

       &       &  71.601 &  71.360 &        \\

       &       &  74.556 &  76.382 &        \\

       &       &  72.134 &  71.404 &        \\

       &       &  75.176 &  76.285 &        \\

990926 & 32653 & 350.600 &  -6.100 &  3.760 \\

       &       & 351.358 &  -9.784 &        \\

       &       & 349.452 &  -2.518 &        \\

       &       & 349.287 &  -9.628 &        \\

       &       & 347.741 &  -3.646 &        \\

991005 & 15265 & 329.000 &  37.900 &  1.840 \\

       &       &   0.000 &   0.000 &        \\

       &       &   0.000 &   0.000 &        \\

       &       &   0.000 &   0.000 &        \\

       &       &   0.000 &   0.000 &        \\

991120 & 27069 & 153.400 & -28.700 &  2.060 \\

       &       & 153.311 & -30.759 &        \\

       &       & 152.564 & -26.777 &        \\

       &       & 153.536 & -30.757 &        \\

       &       & 152.770 & -26.717 &        \\

000102 & 27709 & 213.200 & -13.900 &  2.720 \\

       &       &   0.000 &   0.000 &        \\

       &       &   0.000 &   0.000 &        \\

       &       &   0.000 &   0.000 &        \\

       &       &   0.000 &   0.000 &        \\

000205 & 45486 & 203.000 &  14.600 &  2.000 \\

       &       & 204.808 &  13.637 &        \\

       &       & 201.462 &  15.941 &        \\

       &       & 204.665 &  13.421 &        \\

       &       & 201.297 &  15.740 &        \\

000206 & 09183 &  28.100 &  18.200 &  2.130 \\

       &       &  29.289 &  16.398 &        \\

       &       &  30.195 &  17.451 &        \\

       &       &  27.932 &  16.076 &        \\

       &       &  30.257 &  18.792 &        \\

000403 & 13199 & 275.100 & -10.700 &  1.710 \\

       &       & 272.977 & -10.966 &        \\

       &       & 274.144 & -11.737 &        \\

       &       & 272.758 & -10.720 &        \\

       &       & 273.909 & -11.487 &        \\

000405 & 77386 & 226.900 & -52.500 &  2.640 \\

       &       &   0.000 &   0.000 &        \\

       &       &   0.000 &   0.000 &        \\

       &       &   0.000 &   0.000 &        \\

       &       &   0.000 &   0.000 &        \\

000420 & 61374 & 104.100 &  54.200 &  1.840 \\

       &       & 102.483 &  52.632 &        \\

       &       & 107.083 &  53.652 &        \\

       &       & 101.930 &  52.887 &        \\

       &       & 107.232 &  54.064 &        \\

000511 & 66298 &  48.700 &  38.000 &  2.260 \\

       &       &  47.337 &  36.019 &        \\

       &       &  50.491 &  39.779 &        \\

       &       &  46.089 &  37.093 &        \\

       &       &  48.679 &  40.260 &        \\

\enddata

\end{deluxetable}

\end{document}